\documentclass{spie}
\usepackage[dvips]{graphicx}  
\title{Large Synoptic Survey Telescope: Overview}

\author{J. Anthony Tyson$^a$ and the $LSST$ Collaboration
\skiplinehalf
$^a$Bell Labs, Lucent Technologies, Murray Hill, NJ, USA}

\authorinfo{Further author information: send correspondence to
tyson@bell-labs.com}

\begin{document}
\maketitle

\begin{abstract}
A large wide-field telescope and camera with optical throughput over 
200 m$^2$ deg$^2$ -- a factor of 50 beyond what we currently have -- would enable the detection
of faint moving or bursting optical objects: from Earth threatening asteroids
to energetic events at the edge of the optical universe.
An optimized design for $LSST$ is a 8.4 m telescope with a 3 degree field of view and
an optical throughput of 260 m$^2$ deg$^2$. 
With its large throughput and dedicated all-sky monitoring mode, the $LSST$ 
will reach 24th magnitude in a single 10 second exposure, 
opening unexplored regions of astronomical parameter space.
The heart of the 2.3 Gpixel camera will be an
array of imager modules with 10 $\mu$m pixels.
Once each month $LSST$  will survey up to 14,000 deg$^2$ of the sky with 
many $\sim$10 second exposures.  
Over time $LSST$ will survey 30,000 deg$^2$ deeply in multiple bandpasses,
enabling innovative investigations ranging from galactic structure to
cosmology. This is a shift in paradigm for optical astronomy: from ``survey follow-up''
to ``survey direct science.'' The resulting real-time data products and fifteen
petabyte time-tagged imaging database and
photometric catalog will provide a unique resource.
A collaboration of $\sim$80 engineers and scientists is gearing up to confront
this exciting challenge.
\end{abstract}

\smallskip

\section{Introduction}

Recent progress on large digital surveys like 2MASS and SDSS 
has demonstrated that many astrophysical topics can only be
addressed by exploring wide fields with dedicated telescopes. 
Three recent NAS studies have recommended the Large Synoptic Survey Telescope ($LSST$).
Far more than a telescope, this unique data facility will open new windows on the universe.
Bold explorations, ranging from Earth's vicinity to the edge of the optical
universe, would share the same data.
With an 8.4m primary, a 2.3 Gpixel imager in an f/1.25 beam, the $LSST$ will go {\it faint fast}.
It will reach 24th mag (5$\sigma$) in 10 seconds, and will survey up to 14,000 square degrees
three times per month. Over a period of years, 30,000 square degrees will be surveyed in multiple
bands and the co-added images will go to 27th magnitude\cite{lsst,tyson01}.
These data will be reduced in real time and the resulting images, 
photometric catalogs, and search tools will be public.

This project is enabled by advances in microelectronics, large optics fabrication and metrology,
and software.  Several key programs serve to demonstrate the breadth of science that will emerge
from this facility.  Each is critically dependent on surveying a large volume.
They are all driven by their need to go faint fast, but for different
reasons in each case:

\begin{itemize}

\item{A survey of Near Earth Objects, specifically so-called ``Potentially Hazardous Objects''
which have Earth crossing orbits, down to 250 m in size.  These objects move fast, typically 0.1 
arcsec per second, and must be
detected in multiple $\sim$10 second exposures for linking and orbit
determination.  This is to avoid misidentification
due to confusion with the main belt asteroids.}

\item{Optical Transients to the edge of the optical universe. Current surveys for transients
nibble at the edges of this new window: some are wide-area but limited to bright objects, 
others have pencil beam 
long exposures going faint but losing the short exposure needed for burst or fast moving
object detection.  We need
a large aperture fast optical system which can reach 25th magnitude in less than 20 seconds.}

\item{The equation of state $w$ of dark energy from a large sample of mass clusters back to half
the age of the universe.
300,000 clusters weighed via weak gravitational lens 3-d tomography, using color-redshifts,
could constrain $w$ to one percent. 
Combined with cosmic microwave background data, this will test
the foundations of our cosmology, independent of SN observations.
Going faint fast is necessary to control systematic errors from changing low-level
optical distortions in the telescope.}

\end{itemize}

Curiously, all these projects make use of the same data: short exposures in multiple filters.
Community input has generated a longer list of exciting science which
will result from the multi-color multi-epoch deep imaging which $LSST$ will process and archive:
Kuiper belt objects, intergalactic stars, Galactic structure encompassing the entire local group, QSOs
and AGN, microlensing, rare new objects, and a proper motion survey resulting from unprecedented
astrometry of a dense grid of stars.
These projects have a common thread: the science largely comes from {\it within} the $LSST$
photometric database. This represents a change in paradigm for astronomy and for optical
surveys in particular.

Over the past two years a series of workshops and meetings have refined the science case and, in
turn, the technical requirements for this unique facility.  There is now an $LSST$ Science Working Group,
as well as individual teams addressing software, optics, camera, and operations.
Flow-down to technical specifications for the entire $LSST$ system led to conceptual design
of the f/1.25 wide-field and 7m effective aperture capable of going faint fast.  
The resulting 3-mirror optical design \cite{angel00} has been optimized \cite{seppala02} to
produce crisp aberration free images over the 7 square degree focal plane.

Mirrors and camera optics are manufacturable \cite{sommargren02} and 
the convex secondary has minimal asphere.  The focal plane
is flat or nearly flat, with a radius of curvature over 30 m.  This has important implications for the
focal plane array (FPA) fabrication as well as the FPA module size.
It is generally agreed that the two biggest technical challenges at this point, entering into
Phase-A engineering, are (1) the camera, and (2) software.  The biggest challenge in the camera
is the FPA module technology.  It is necessary to take a systems approach to this project, since the various
components of the system (optics, detector, telescope, data, software, data products) are 
strongly coupled.

\section{The Telescope}

\subsection{Optics}
 
Modern $\sim$ 8 m aperture telescopes are not
only larger than the previous 4 m class, they yield sharper images.  The
advances that have made this possible are the elimination of local
convection and imperfections in the mirrors and their alignment.  The
residual higher altitude turbulence limits images to a median of $\sim$ 0.5
arcsec, with occasional periods of images as sharp as 0.4 arcsec or
better. 

\begin{figure}[h]                                               
\begin{center}
\includegraphics[height=4in]{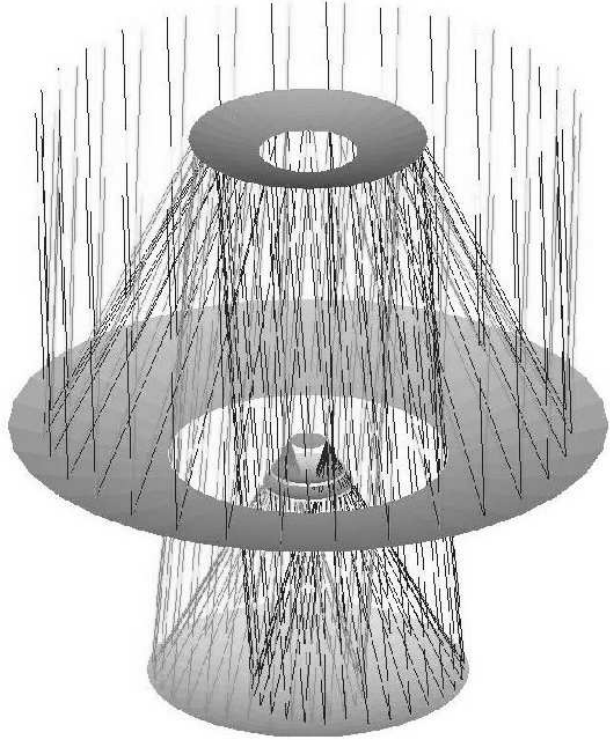}
\end{center}
\caption{\small \setlength{\baselineskip}{10pt}
Optical layout with rays at $\pm$ 1.5$^{\circ}$ field angle.
Dimensions are: primary 8.4m, secondary 3.5m, and tertiary 4.2m.
With five aspheric surfaces to adjust, low aberration images of less 
than 0.2 arcsec (80\% energy) are obtained over the entire 7 square degree
focal plane. This was done while maintaining minimum asphericity for the 
convex secondary for manufacturability.}
\label{rays}
\end{figure}

The strength of $LSST$ lies not in a small diffraction limited field of view but in
having the widest possible field with the largest aperture.  Equipped
with a large mosaic of detectors capable of reaching sensitivity limits set only
by sky photon noise and atmospheric seeing, such a telescope has
unique scientific potential.  
For the dual requirements of large field and low aberration PSFs, 
optical systems based on one or two mirrors are inadequate, and
three mirror systems with strongly aspheric primary and secondary must
be used.
Three-mirror telescopes 
with a parabolic primary, convex spherical secondary and a
concave spherical tertiary of equal but opposite curvature,  
were first explored by Paul \cite{paul35}. 
The image
is formed midway between secondary and tertiary, with good correction
over a wide field.  
The secondary, located at the center of curvature of the
tertiary, has added correction for spherical aberration.
For $LSST$ general 3-mirror systems were explored, allowing all three
mirrors as well as two inner surfaces of camera windows
to have an aspheric figure optimized for wide field.
The focal surface was not constrained to be flat, since the dedicated
detector mosaics can be configured to approximate a curved field, with
each flat device tangent to the focal surface.
The optimized design has the tertiary mirror located
significantly behind the primary (Figure~\ref{rays}).  

The advantages of this
arrangement for wide field imaging were recognized by Willstrop\cite{willstrop84}.
In this recent 5-surface optimization the effects of longitudinal and chromatic
aberration are almost eliminated.  
Detector obscuration is minimized by making the primary and secondary
together nearly afocal. The 3 deg field is
completely baffled against stray sky light illumination with total
obscuration and vignetting held to 26\% at the field center and rising
to 38\% at the field edge.
Recently the design has been optimized to deliver high quality images
over the entire focal plane \cite{seppala02}.
 
There are many issues to consider in the design of such a facility:
what are the trades involved in making the aperture smaller or splitting the
effective aperture into multiple telescopes. The optimal strategy hinges on
two factors: the optical throughput per telescope/camera and the constraints on
the exposure time due to science requirements or the necessity of reaching
the sky noise limit.

\subsubsection{The power of A$\Omega$}

Neglecting exposure time requirements, a general measure of the survey 
efficiency of a telescope is the figure of merit
$A\Omega\epsilon/d\Omega$.  Here A is the collecting area, $\Omega$ the
solid angle of the field of view, $\epsilon$ the overall efficiency including
detector quantum
efficiency and d$\Omega$ the solid angle of the seeing limited image.
In given integration time the size of field larger than $\Omega$ that
can be explored to given depth is directly proportional to this figure
of merit.  In the sky noise limit, the surveyed area of sky $\Omega_{\mathrm {survey}}$
to a limiting flux $F_{\mathrm {obj}}$ per unit time to some S/N is proportional
to A$\Omega$:
 

$${\Omega_{\mathrm {survey}} \over t} \sim {{F_\mathrm {obj}^2 A \Omega \epsilon} \over {{(S/N)^2} F_{\mathrm {sky}} \delta\Omega}}$$


\noindent
where $F_{\mathrm {sky}}$ is the sky background flux and $\delta\Omega$ is the
PSF footprint.

\begin{figure}[h]
\begin{center}
\includegraphics[height=3in]{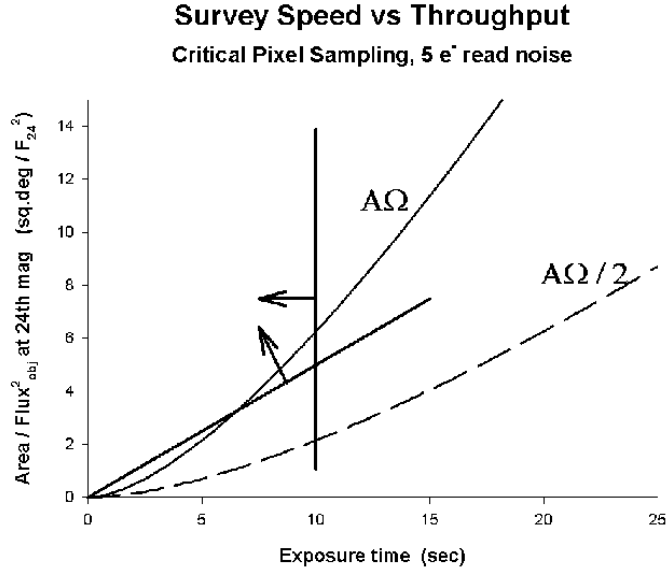}
\end{center}
\caption{\small \setlength{\baselineskip}{10pt} This plot shows the
effect of telescope plus camera optical throughput on the survey efficiency
as a function of exposure time, for a given detector noise. Two
fiducial throughputs A$\Omega$ are shown, differing by a factor of
two. The solid line represents $LSST$ at A$\Omega$=265 m$^2$
deg$^2$. A 2.3 Gpixel camera (critical sampling) is assumed. The
ordinate is the area surveyed to a given depth per exposure,
normalized to 24th V mag.  Survey speed increases with A$\Omega$.  But
for the low throughput telescope the sky noise limit is never reached
for exposure times less that the maximum (10 sec as shown) set by the
science requirements. Other science drivers require deep coverage of
the entire visible sky every few days (upper left region), giving a
minimum data rate.}
\label{texp}
\end{figure}

Note that this equation is independent of the exposure time.  This is because sky noise
limited performance was assumed. For the $LSST$ science
drivers the exposure time needs to be less than 20 sec. 
While sky limited operation is not critical for bright
objects, the regime of $LSST$ imaging is the faint limit. The square root of
the sky background counts in an exposure must greatly exceed the noise
of the device.  This creates a critical throughput below which the telescope-camera
cannot operate efficiently as shown in Figure~\ref{texp}.
For a given number of pixels in the FPA, $n_{\mathrm {pix}}$, the flux into each pixel is
$F_{\mathrm {sky}} A \Omega / n_{\mathrm {pix}}$ and the sky noise limit is reached when
$t_{\mathrm {exp}} F_{\mathrm {sky}} A \Omega \epsilon / n_{\mathrm {pix}}~ > ~ N^2_{\mathrm {device}}$.
In Figure~\ref{texp}, the solid curve represents $LSST$ at A$\Omega$=265 m$^2$ deg$^2$.
Two science requirements are shown: short exposures (optical transients and non-trailing
for moving objects) $t<10$s, and sufficient survey pace in deg$^2$ per night in order
to cover the entire visible sky in 3 nights (NEO detection).
Surveys constrained to short exposures get a double boost from high A$\Omega$: once from
area covered to a limiting flux, and once again from shorter times (and thus faster
survey pace) required to overcome device noise.

For example, if the detector modules are 2k$\times$2k and must be read
out in 2 sec, then a read+thermal noise of 5-10 e$^-$ is typical of
current CMOS or CCD devices cooled to -40 C.  In a ten-second exposure we must therefore
build up at least 1000 e$^-$ of sky in order to have less than a 20\%
decrease in S/N ratio due to read noise.  This leads to a critical
A$\Omega$ which must be exceeded for each telescope in an array or for
a single telescope, in order to go faint fast.  The properties of the
detector become important for such short exposures.  If the exposure
is too short, CCD traps are not filled and/or the detector (whether
CMOS or CCD) is not sky-noise limited.  Low f-ratio illumination is
crucial.  

\subsubsection{Tolerance for maintenance of image quality}
 
The ability to go faint fast is good for control of image systematics.
Weak lensing places demands on the image aberrations over the field
and their stability.  With current telescopes where this is not controlled, 
on-line convolution of the PSF, fit
to all the stars over the field in a given exposure, can correct for
PSF ellipticities as bad as 10\%, reducing
the PSF ellipticity to an average value of about 0.1\%.  
For the low-level cosmic shear work we will need to fit the 
PSF to many stars over a large field and reduce the PSF ellipticity to
0.01\%. However, the mass cluster 3-D tomographic mode, in which typical
clusters generate shears of 0.1, will not require such
low level shear systematics control.  
This increased dynamic range for PSF correction will be
accommodated by a combination of hardware and software PSF control.  We
have found that in single exposures reaching 24th mag that there will be
enough stars per unit area in the field to make a robust software deconvolution of
the spatially varying PSF, if the PSF ellipticity can be kept below
5\%.  

The $LSST$ is sufficiently fast that this magnitude limit can be
reached in only a 10 second exposure.  
To enable this software AO, the delivered
image quality of the $LSST$ must be controlled.  Thus, we require
hardware control of the telescope optics at that level, along with the
corresponding metrology.
The $LSST$ Collaboration have undertaken a realistic study of the image
aberrations under varying amounts of primary mirror decenter and
correcting tilt/decenter of the secondary mirror.  The results are
quite encouraging \cite{claver02}.  
A finite element analysis of the rigid support for the primary
suggests that decenters of the primary larger than 100 microns would
be unlikely. 
In order to tie
all the optical elements of the telescope together, off-the-shelf
laser metrology (20 micron tolerance), along with four stepped out of focus
detectors at the field edge, will be sufficient.

Finally, there will be hundreds of exposures of each 7 deg$^2$ part
of the sky with the camera rotated at various orientations, thus averaging
over remaining systematics and affording the opportunity to cut the
sample in various ways.  One obvious way is to discard individual exposures
with substandard PSF, co-adding images with only the best PSF. 
Selecting the best 50\% seeing, FWHM of 0.5 arcsec or better should be achieved
in the co-added stack. How will the residual variations in the telescope optics 
PSF affect low-level gravitational shear measurements? We have done tests in 
which we convolve slightly sheared multiple
orientations of the HST HDF fields with the telescope PSF at various places
in the focal plane and with a 0.5 arcsec atmosphere PSF. We find a negligible 
effect for the range of telescope PSF variations which go uncorrected: shear
error is dominated by the soruce galaxy ellipticity noise rather than telescope
systematics.  

To maximize the figure of merit for an imaging telescope, it should be
in a site with excellent seeing and its focal length must be chosen so
the detector pixels will adequately sample seeing limited images.
Experience shows that good telescopes at the best sites will deliver
images of 0.4 - 0.5 arcsec often.
The pixel sampling should be no worse than 0.2 arcsec (the
Nyquist sampling criterion) to avoid further significant image
degradation, thus each square degree on the sky must be sampled
by over 320 million pixels in the detector mosaic.

\subsection{Comparison with some existing and proposed imaging telescopes}
 
\begin{figure}[h]
\begin{center}
\includegraphics[height=4in]{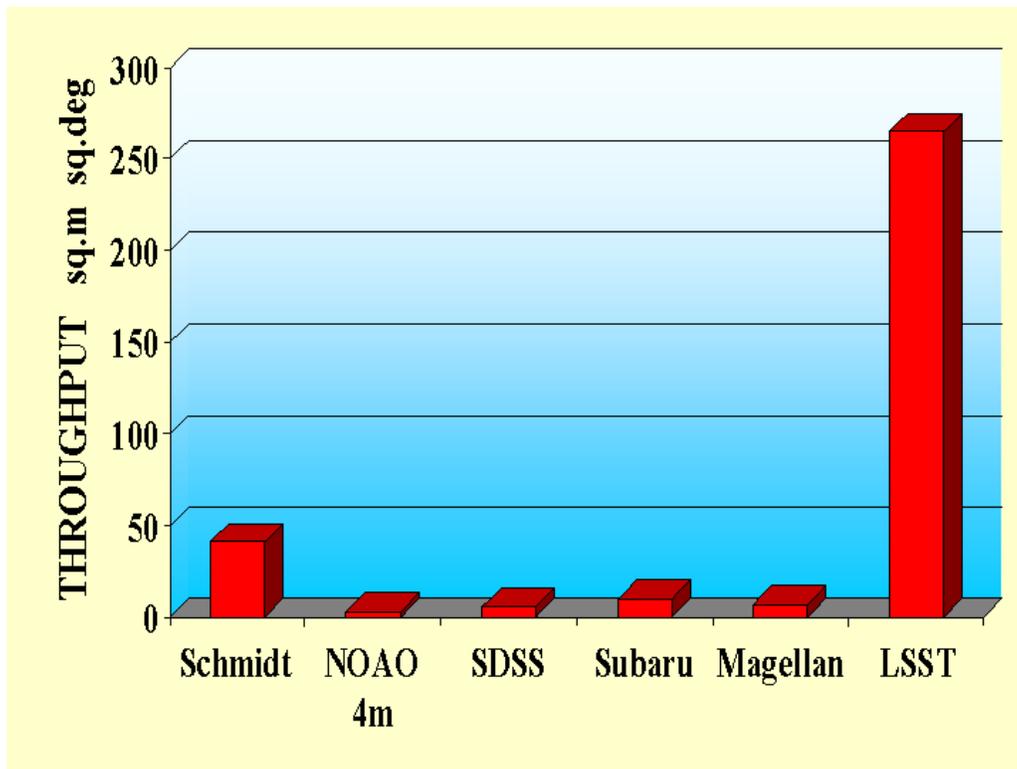}
\end{center}
\caption{\small \setlength{\baselineskip}{10pt}
The optical throughput of various survey telescopes
is compared. While the old Schmidt has moderately high throughput, its delivered
image quality is poor. The science requirement of going faint fast leads to
the need for a telescope/camera with throughput over 200 m$^2$ deg$^2$ and
with sky background limited exposures of 10 seconds. 
}
\label{throughputs}
\end{figure}

Comparing the 8.4 m telescope with the SDSS, and allowing also for its
increased pixel sampling and resolution, the advantage in figure of
merit is by a factor of close to 200.  The wide field optical cameras
to be used with existing new-technology larger telescopes 
have throughputs which are not substantially larger than the
SDSS, in the range 5 - 10 m$^2$ deg$^2$.  
The science drivers for
the $LSST$ put a premium on large $A\Omega$.  The throughput of various
telescopes useful for good image quality wide-field work is compared
in Figure~\ref{throughputs}.
This $LSST$ facility will provide a capability that is completely
beyond any existing telescope and uncovers a region of parameter
space orders of magnitude beyond current limitations. 
Capable of reaching 24th mag in a single ten second exposure,
the $LSST$ will
be able to reach 10$\sigma$ limiting magnitudes of $U=26.7, B=27.8,
V=27.9, R=27.6$, and $I=26.8$ over a 7 deg$^2$ field in 3 nights of
dark time.

\section{The Camera}

\begin{figure}[h]                                               
\begin{center}
\includegraphics[height=4in]{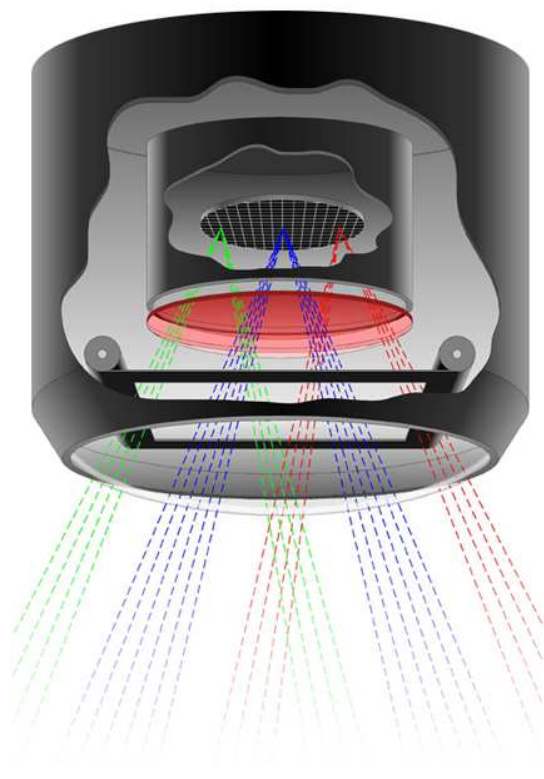}
\end{center}
\caption{\small \setlength{\baselineskip}{10pt}
The Xenon filled inner dewar of the $LSST$ camera protects the detector
array, which is cooled only to -30 to -40 C. Outside this dewar is a dry nitrogen
filled can housing the filter and shutter
mechanisms.  Four filters can be on storage mechanism -- only
one filter is shown in place in this drawing.
}
\label{dmt-dewar}
\end{figure}

A plate scale of  51$\mu$m/arcsec ensures that seeing-limited images
will be well sampled by the 10 $\mu$m pixels. 
The optimized
Paul optical design puts the camera in the beam from the secondary:
not a lot of room for filter and shutter mechanisms.
For a large aperture $A$ the resulting detector array is necessarily large,
and the usual vacuum dewar would have an unusually large and thick glass
window. This window would have to be curved to take the 14-ton load,
and mechanisms like a shutter and filter changer which should reside outside 
the vacuum would be huge and would be in the beam.  
However, the large throughput results in very short exposure times (sky noise
limit is reached within seconds).  Thus, the detector need not be cooled
below -40 deg C, and the dewar may be filled with a low thermal conductivity
gas like Xenon.  Figure~\ref{dmt-dewar} shows this layout.  The
dewar window is thin and not as curved, and there is room in the outer
can (filled with dry nitrogen also at atmospheric pressure) for the 
required mechanical assemblies.

Ghosts from multiple reflections of bright stars can be a problem in
wide-field telescopes, particularly with multiple refractive element
correctors. The $LSST$ is designed to avoid such a corrector,
and the dewar window and filter are placed so far from the CCD that,
with standard broadband AR coats, they will supply insignificant
ghosting from the inevitable 5th mag stars. 
The dewar window is 1.3 m diameter, but need not take vacuum loading
since the dewar will be filled with atmospheric pressure
low-conductivity gas.  Its convex meniscus shape adds shell stiffness.
 

In the current $LSST$ design the FPA is 55 cm in diameter. 
Cooling requirements are not severe for the detector module mosaic.  The criterion
is that the dark rate be less than the sky photon rate in the darkest filters.  
With an MPP
CCD or current CMOS devices, low dark rates can be achieved at a device
temperature of -30 to -40 C.  In the worst case of
a 10 second exposure in a dark band a read noise of $\sim$ 4 e$^-$ rms
will be acceptable.
 

\subsection{Mechanical issues}

\begin{figure}
\begin{center}
\includegraphics[height=5in]{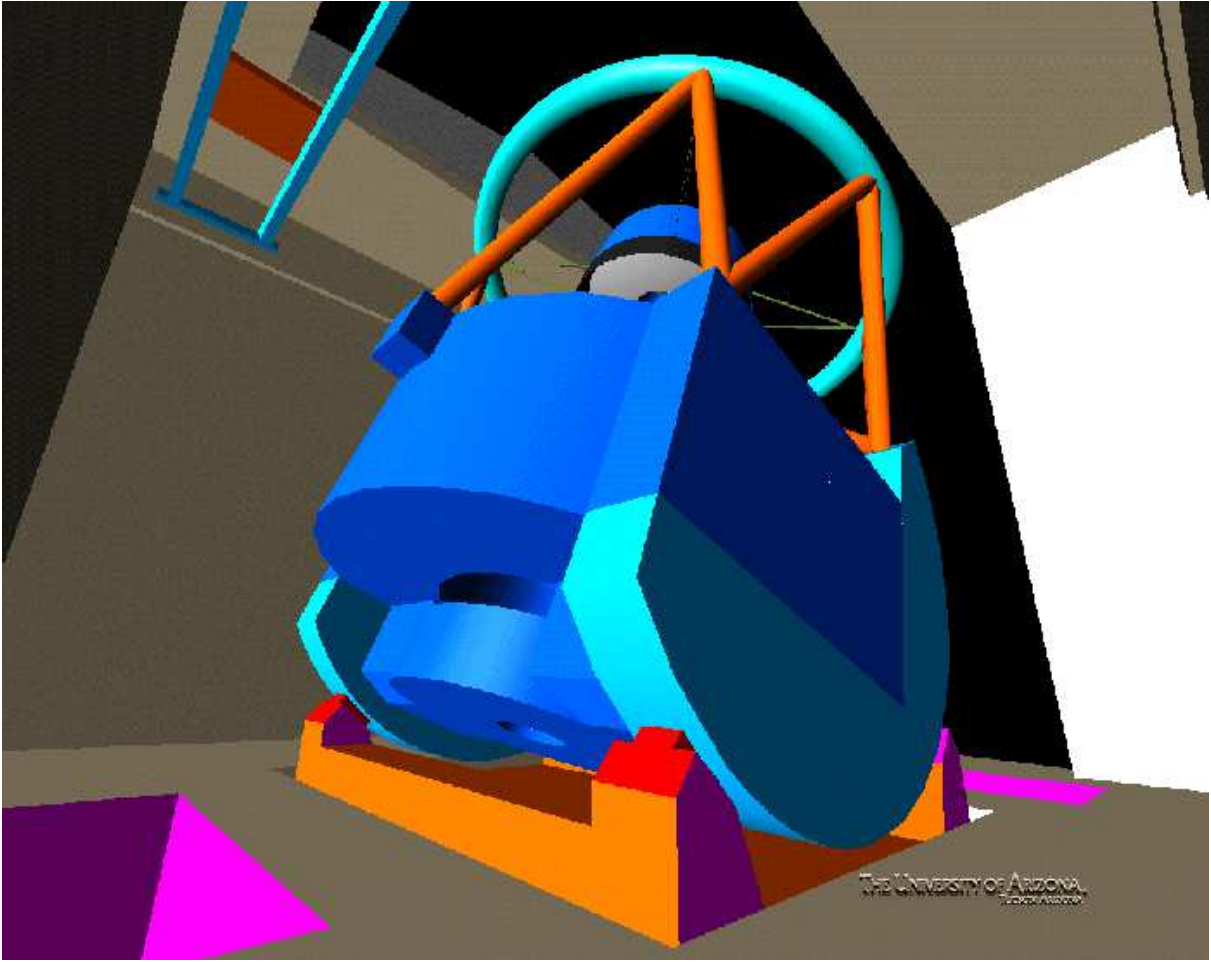}
\end{center}
\caption{\small \setlength{\baselineskip}{10pt}
A stiff compact design capable of fast slewing enables $LSST$ to fit
inside a small dome no larger than Magellan's. (Davison 2000)
}
\label{lsst-wd}
\end{figure}

While many of the exposures are shifted on the sky by a fraction
of the 3 deg field, it will be necessary to slew the telescope by
about three degrees fairly often. With a 2 second read time, it will
be important to design a system which can repoint gracefully and fast.
Exposures of 10 sec do not require guide stars if the tracking
is sufficiently accurate.
Slew and settle time for the telescope are dependent on overall structural
stiffness.  The current design by Davison\cite{davison02} (Figure \ref{lsst-wd}) 
is compact and stiff, with
a first resonance above 8 Hz.  A related issue is the servo system,
and an end-to-end systems analysis will be undertaken. 

There are two main mechanical challenges with the camera.
As seen in Figure~\ref{rays}, the camera is in the beam from the secondary.
The axial cross section of the camera assembly is thus constrained, prohibiting
conventional outboard filter changing mechanisms and shutter mechanics.
Whether a shutter is even necessary depends on the nature of the detector
module.  CMOS arrays, having several transistors associated with each
pixel, are self-shuttering.  CCDs have been developed in which a
horizontal electric field can pinch off charge collection, but this fails for
red wavelengths where photons penetrate 50 microns. So CCD arrays will
require some type of shutter.  
Engineers at LLNL have been studying several compact shutter
designs which promise long mean time to failure together with uniform
illumination over the focal plane. While one of the desirable properties
of the detector module is self-shuttering, 
either CMOS or CCD solutions to the $LSST$ detector modules are viable.

The filter change mechanism is the other challenge. The science drivers, even for
the NEO survey, require several band exposures.  Some science can settle for
long filter change times, while some projects require that at least a few
filters be available within minutes.  Compact filter change/storage designs
have been studied for the camera.  In particular, a solution is now
being studied in which four filters are stored in the outer chamber
of the dewar \cite{starr02}. A filter cassette design, with storage behind the
primary mirror is also being studied.

\section{The Focal Plane Array}

The design of the $LSST$ started with the output: the question was
``what data are required for the science?''  This drives the number of
pixels, their response vs wavelength, and their dark and read noise.  
Covering the 3$^\circ$ diameter field of view with 0.2 arcsec
pixels (to provide critical sampling in the best seeing) requires 2.3
Gpix.  

The size of the pixels is set by a number of considerations:
(1) the minimum and maximum acceptable stored charge (dynamic range per pixel),
(2) the requirement that we maximize the $A\Omega$ product,
(3) sky noise limited operation with 10 sec exposures,
(4) depth of focus (a consideration at near-IR where photons penetrate silicon),
(5) maximize the near-IR QE (a pixel geometry constraint),
(6) good performance in the near-IR, and (7) minimize device area in order to
maximize yield.  




These demands jointly constrain the pixel size to about 10 microns.
The $LSST$ camera will use state-of-the-art imager and microelectronics technology which was
developed over the past few years for a variety of applications ranging from HDTV to fast low-noise
multiplexers for IR arrays to fast low noise readout arrays.
High QE in the near-IR favors $>30$ micron thick fully depleted high-resistivity
silicon.  Recently, spectacular QE curves have been obtained for such
devices by several labs.
The individual array modules will be small, 1K or perhaps 2K, so that
the the circular field and the curvature (if any) of the focal plane can be
matched precisely: the depth of focus is 10$\mu$m.  
Parallel multiplexing many discrete modules also allows for fast readout,
which will be critical for efficiency. The clocking electronics will
be integrated with the individual detectors and there are several
attractive options for analog and digital ULSI packaging which
minimize the interconnections. 


There are several factors leading to the need to break up the FPA into small modules.  
The worst case curved focal plane accommodates devices smaller than 2 cm, without bending.  
Device yield
decreases as the area increases, leading to practical FPA segments of several cm or smaller size.
The readout time
requirement, given that the exposures are as short as 10 sec, is 2 sec.  With a pixel size of 10 microns,
low noise ($< 5$e) reads of CCDs can be performed with correlated double sampling at
individual device rates of less than 1 Mpix/sec, leading to a size limit of roughly 2 Mpix or 1.5 cm.
CMOS arrays can read faster.

We require an FPA with imager modules with the following properties:
4-side edge buttable\cite{lesser02}, overall 90\% fill factor or higher, high QE 350-1050 nm, and integrated 
readout and A/D electronics. The idea is to develop self-contained imager modules which have a
DC voltage in, bi-directional communication (possibly FPGAs), and digital data out.  
For the $LSST$ imager we are examining several CCD technologies as well as CMOS arrays\cite{bai00}.
Another possibility, already under development, is a module consisting
of an array of CCDs, with integrated controller.  These matters
are discussed in a companion paper by Lesser and Tyson \cite{lesser-tyson02}.

\section{Observing Strategy}

While each of the science programs use the multiple sky-limited short exposures
in different ways, the observing strategy is set mostly by the transient and moving object survey programs.
The number of Potentially Hazardous Asteroids (PHAs) rises steeply as their size decreases, and
they are extremely faint.  No current telescopes are capable of
surveying for them. 
Deep imaging is necessary but not sufficient.
In oder to avoid confusion with main belt asteroids one must go faint fast.
Preliminary orbits must be reliably linked across nights and weeks.
A PHA moving with an angular speed of up to 2.5 deg/day
will not trail in 2x10 sec exposures, and thus it will be
detected in the same way as the main belt asteroids (whose
typical velocity is 0.3 deg/day). 
For R$<$24 mag) the
typical distance between two main belt asteroids is 2 arcmin on the Ecliptic. 
They can be
linked in two 10 sec exposures 15 min apart, in which they will move up to
10 arcsec. With this 15 min based velocity vector they can
be linked again within 3 nights with false positive rate of
less than 1\%. 
The interesting range for PHAs extends up to $\sim$10 deg/day 
and PHAs with angular speeds greater than 2.5 deg/day 
will be slightly trailed.
The trailing will help to link them in two exposures taken 15
minutes apart, but also increases the detectable size limit.

%
%
%
%
%
%
%
%

\section{Data Rate and Pipeline}

The $LSST$ facility will produce over 7 TB of raw data per night, compressed.
Such data rates are commonplace in radio astronomy and particle physics.
The huge difference here is that we must keep all this data as well as
analyze it in real time!
The required hardware for image processing, including frame dewarping,
is achievable based on what is currently under development in the lab.
Routine image processing on
large images is easily parallelized and ideally suited to clusters of
inexpensive, commodity computers, especially given the parallel nature
of the readout.  
The HDTV and other industries are driving the development of 
storage capability which will become available at reasonable cost in
just the next few years.
Disk access will not be a limiting factor either. 
The software pipelines will present more of a challenge. 
Even here, however, previous
projects such as the searches optical transients have
demonstrated that efficient real-time use can be made of very large
data streams. These projects,like SuperMACHO and the Deep Lens Survey,
had to develop pipeline software for fast transient detection\cite{smith02,cook02,wittman02}.  
Likewise, the SDSS had to develop a photometric pipeline\cite{lupton02}.
These surveys are precursors to the $LSST$, and the lessons learned and
many of the algorithms will be put to work on the $LSST$ pipelines.

The $LSST$ photometric pipeline will process the data through flat
fielding and will also produce a running catalog of detected and
photometered objects.  There will be various taps in the pipeline
for data quality assessment at various stages of processing.
Some aspects of this pipeline will have to be detector specific.
Hopefully, we will gain insight into the detector features before the
main camera is completed.
The architecture of the $LSST$ data flow is still being worked out.
It is already known that after the automated photometric pipeline
there will be several parallel pipelines producing different data products.
One of them will be the Transient pipeline, tuned to detect transient
and also moving objects.  There will also be a parallel pipeline 
for Moving Objects which will take inputs from the database, the Transient
pipeline, and the camera. All these pipelines will most likely be evolutions
of similar ones which are now being refined for current surveys.

The job of these transient pipelines will be aided immensely by having a
pre-existing deep multicolor photometric catalog of the area being surveyed.
An initial 2 years may be used for just this purpose, building up depth
as the main $LSST$ survey proceeds later.  One of the lessons we have learned
from previous surveys is that the cost of the software effort will be
comparable to the telescope hardware or camera.  With over 10 TB of new data
every night, we have to do it right.  Cutting corners on software support
is not cost-effective.

More important than how we
process the data is what we do with it.  The traditional mode of
observing must give way to a dedicated program, with the community
sharing the data.  In addition to weak lensing, the search for
near-Earth objects, outer solar system objects, and high-redshift SN
will benefit greatly, and a vast amount of parameter space will be
opened up in the search for GRB counterparts and previously unknown
types of optical transients.  These projects can {\it only} be
done with a telescope that can go deep over a wide field in a short
time, combined with sensible data management and analysis.

\section{Community access, data analysis and mining}

All the $LSST$ will be public, the transient and moving objects immediately and
co-added imaging and photometric catalogs as soon as calibrated and quality assessed.
Data products will consist of photometric
catalogs which will be continuously updating during the survey, optical
burst announcements and characteristics, a moving object database,
deep co-added images in at least 5 bands (updated on a regular schedule),
and the huge time-tagged processed image database which will climb to
around 15 Petabytes after ten years.

Alerts and some catalogs could easily be on-line.  The full photometric
catalog will probably be too big for easy remote access. This of course
is also true of the Petabyte imaging database.  These various data form
a pyramid. Most of the users will start from the top, and there must be
tools for searching remotely.  But large users must be accommodated in
a different way.  Most of the science to emerge from the $LSST$ and its
database will be statistical in nature and will thus involve correlations
over large parts of the data.  In fact it is often the unexpected correlation
that is the most interesting scientifically.  Two things must happen 
to enable this key mode of data analysis: efficient algorithms for
statistical analysis must be developed, and this computation must be via
high bandwidth to the spinning disk database storage facilities.  Data
backup via multiple nodes would thus serve two purposes.
Clusters of computers co-located with the disk farms would assure
both community access and high local I/O bandwidth.

Crafting the software pipelines and developing efficient database management
tools and the algorithms for data mining will be more challenging than the
pre-processing computation.  Most algorithms scale as a power of the number of
objects.  This traditional approach will be impossible here.  Approximate
statistical techniques, where the approximation is within the sample
variance, and with non-polynomial scaling, will be key \cite{szapudi00,moore01,szalay02}.
Quality assessment in all phases of the data chain will be important, including
the tests of the photometric catalogs and the imaging database.
Low level systematic errors must be found and quantified.
This process, and the direct involvement of university scientists, 
could be motivated by funding a key science result.


\begin{thebibliography}{99}

\bibitem{angel00} Angel, J.~R.~P., Lesser, M., Sarlot, R., \& Dunham, T. 
``Design for an 8 m Telescope to Image a 3 Degree Field at f/1.25 - the
Dark Matter Telescope'' 
in {\it Imaging the Universe in Three Dimensions: Astrophysics with Advanced
Imaging Devices}, eds. W. van Breugel, J. Bland-Hawthorne,
ASP Conference Series 195, (Astron. Soc. Pacific, San Francisco) 81 (2000).

\bibitem{bai00} Y. Bai, J.~T. Montroy, J.~D. Blackwell, M.~C. Farris,
L.~J. Kozlowski, and K. Vural,  
``Development of hybrid CMOS visible focal plane arrays at Rockwell'',
Proc. SPIE Vol. 4028, p. 174-182 (2000).

\bibitem{claver02} C.~F. Claver and J.~H. Burge, ``Active Alignment of the 3-Mirror LSST,''
Proc. SPIE 4836-25 (2002).

\bibitem{cook02} K. Cook, ``Exploring the Time Domain with LSST: Lessons learned from MACHO''
Proc. SPIE 4836-52 (2002).

\bibitem{davison02} W. Davison, ``Large Synoptic Survey Telescope mechanical structure
and design'' Proc. SPIE 4836-18 (2002).


\bibitem{lesser02} M.~P. Lesser and D. Ouellette, ``Fully Buttable Imagers,'' 
{\it Scientific Detectors Workshop 2002}, P. Amico and J. Beletic, eds., (2002) in press.

\bibitem{lesser-tyson02} M.~P. Lesser and J.~A. Tyson, ``Focal Plane Technologies for LSST''
Proc. SPIE 4836-38 (2002).

\bibitem{lupton02} R. Lupton and Z. Ivezick, ``SDSS Imaging Pipelines'' Proc. SPIE 4836-49 (2002).

\bibitem{moore01} A. Moore, etal. ``Fast Algorithms and Efficient Statistics: N-Point
Correlation Functions'' in {\it Mining the Sky} Proc. MPA/ESO/MPE Workshop.
eds. A.J. Banday, S. Zaroubi, and M Bartelmann (Springer-Verlag, 2001) p. 71;  astro-ph/0012333 (2001). 

\bibitem{starr02} B.~M. Starr, etal. ``LSST Instrument Concept,'' Proc. SPIE 4836-37 (2002).

\bibitem{szalay02} A. Szalay, ``Petabyte-scale Data Mining: Dream or Reality?''
Proc. SPIE 4836-47 (2002).

\bibitem{szapudi00} I. Szapudi, et al. ``Fast CMB Analyses via Correlation Functions''
astro-ph/0010256 (2000). 




\bibitem{paul35} M. Paul, ``Syst\`{e}m Correcteurs pour R\'{e}flecteurs Astronomiques,'' 
Rev. d'Optique, 14, 169-202 (1935).

\bibitem{willstrop84}  R.~V. Willstrop, 1984 
"The Mersenne-Schmidt -- A three-mirror telescope"  
MNRAS, 210, 597 





\bibitem{lsst} $LSST$ website: http://lsst.org


\bibitem{seppala02} L.~G. Seppala, ``Improved Optical Design for the Large Synoptic
Survey Telescope,''  Proc. SPIE 4836-19 (2002).


\bibitem{smith02} C. Smith, ``SuperMACHO and Supernova Projects at NOAO/CTIO'' Proc. SPIE
4836-55 (2002).

\bibitem{sommargren02} G.~E. Sommargren and J.~H. Burge, ``Testing the Large Synoptic Survey
Telescope (LSST) Aspheric Convex Secondary'' Proc.  SPIE 4836-20 (2002).


\bibitem{tyson01} J.~A. Tyson, \& R. Angel, ``The Large-aperture Synoptic Survey
Telescope'' {\it The New Era of Wide Field Astronomy},
ASP Conference Series, R. Clowes, ed., 232, 347 (2001).

\bibitem{wittman02} D. Wittman, ``Deep Lens Survey'' Proc. SPIE 4836-12 (2002).
\end{thebibliography}
\end{document}